\documentclass[aps,prb,twocolumn,groupedaddress]{revtex4-1}
\usepackage{graphicx}

\begin{document}


\title{Point contact spectroscopy in the superconducting and normal state of $\mathrm{NaFe_{1-\textit{x}}Co_\textit{x}As}$}


\author{H. Z. Arham$^1$}
\email[E-mail: ]{arham1@illinois.edu}
\author{D. E. Bugaris,$^2$ D. Y. Chung,$^2$ M. G. Kanatzidis$^2$}
\author{L. H. Greene$^1$}

\affiliation{\\\textsuperscript{$1$}Department of Physics and the Frederick Seitz Material Research Laboratory, University of Illinois at Urbana-Champaign, Urbana, Illinois 61801, USA
\\\textsuperscript{$2$}Materials Science Division, Argonne National Laboratory, Argonne, IL 60439, USA}


\date{\today}

\begin{abstract}
We use point contact spectroscopy to probe the superconducting and normal state properties of the iron-based superconductor $\rm{NaFe_{1-\textit{x}}Co_{\textit{x}}As}$ with $\rm{\textit{x} = 0, 0.02, 0.06}$. Andreev spectra corresponding to multiple superconducting gaps are detected in the superconducting phase. For $\rm{\textit{x} = 0.02}$, a broad conductance enhancement around zero bias voltage is detected in both the normal and the superconducting phase. Such a feature is not present in the $\rm{\textit{x} = 0.06}$ samples. We suspect that this enhancement is caused by orbital fluctuations, as previously detected in underdoped $\rm{Ba(Fe_{1-\textit{x}}Co_\textit{x})_2As_2}$ (Phys. Rev. B 85, 214515 (2012)). Occasionally, the superconducting phase shows a distinct asymmetric conductance feature instead of Andreev reflection. We discuss the possible origins of this feature. NaFeAs (the parent compound) grown by two different techniques is probed. Melt-grown NaFeAs shows a normal state conductance enhancement. On the other hand, at low temperatures, flux-grown NaFeAs shows a sharp dip in the conductance at zero bias voltage. The compounds are very reactive in air and the different spectra are likely a reflection of their different oxidation and purity levels.
\end{abstract}


\maketitle

\section{Introduction}

Point contact spectroscopy (PCS) is performed by measuring the differential conductance $dI/dV$ across a metallic junction. When the junction is comprised of a normal metal and a superconductor, the transport is dominated by Andreev reflection.\cite{Andreev} Thus PCS proves to be an extremely useful spectroscopic technique for studying unconventional superconductors. The measured $dI/dV$ curves are sensitive to the magnitude and symmetry of the superconducting order parameter and the spectra may be fit to the Blonder-Tinkham-Klapwijk model (BTK) to obtain this information.\cite{Blonder} PCS was instrumental in determining the precise location of the line nodes for the heavy fermion compound $\rm{CeCoIn_5}$, and in providing direct evidence for the multi-gap nature of the superconductor $\rm{MgB_2}$. \cite{WKPark, MgB2} 

Aside from superconductors, PCS has also proven to be useful in studying compounds with strong electron correlations. In certain heavy fermion compounds, PCS picks up the the hybridization gap and the onset of the Kondo lattice as a Fano lineshape. \cite{WKPark, WKPark2, YifengYang} 

We have previously studied the 122 family of the iron-based superconductors and found evidence for multiple superconducting gaps in their superconducting state and indications of orbital fluctuations in their normal state. \cite{Arham, Arham2, OurReview} 

In this paper we present PCS spectra on the 111 family of the iron-based superconductors, $\rm{NaFe_{1-\textit{x}}Co_{\textit{x}}As}$ with $\rm{\textit{x} = 0, 0.02, 0.06}$. The Andreev spectra for $\rm{\textit{x} = 0.02, 0.06}$ provides evidence for multiple superconducting gaps. We fit our lowest temperature data using the extended BTK model with two s-wave superconducting gaps. \cite{Brinkman} $\rm{NaFe_{0.98}Co_{0.02}As}$ shows a broad enhancement around zero bias voltage that coexists with the Andreev reflection and survives well into the normal state. Such an enhancement does not appear to be present for $\rm{NaFe_{0.94}Co_{0.06}As}$. This enhancement may be indicative of the presence of orbital fluctuations in the normal state of $\rm{NaFe_{0.98}Co_{0.02}As}$. PCS has previously detected orbital fluctuations in underdoped $\rm{Ba(Fe_{1-\textit{x}}Co_\textit{x})_2As_2}$. \cite{Arham} Occasionally instead of Andreev reflection, a distinct asymmetric conductance feature is detected in the superconducting phase. We discuss the possible origins of this feature. Melt-grown NaFeAs shows a conductance enhancement in the normal state while flux-grown NaFeAs shows a sharp dip at zero bias voltage at low temperatures. We consider possible explanations for these behaviors.  

\section{Experimental Results}

Single crystals of $\rm{NaFe_{1-\textit{x}}Co_{\textit{x}}As}$ $\rm{(\textit{x} = 0, 0.02, 0.06)}$ were grown from NaAs flux. Handling of all materials was performed in an Ar-filled glovebox. The starting material $\rm{Fe_{1-\textit{x}}Co_\textit{x}As}$ was prepared from an elemental mixture annealed at temperatures of $\rm{650^\circ}$C, $\rm{700^\circ}$C, $\rm{800^\circ}$C, and $\rm{900^\circ}$C, with intermittent grinding. Phase purity was confirmed by powder X-ray diffraction using a PANalytical X'Pert Pro diffractometer with a Cu K$\alpha$ source operating at 45 kV and 40 mA in a continuous scanning method with a 2$\theta$ range of $\rm{20-90^\circ}$. NaAs was prepared from a mixture of Na and As heated at $\rm{500^\circ}$C. 

The growth of $\rm{NaFe_{1-\textit{x}}Co_{\textit{x}}As}$ single crystals utilized a 6:1 ratio of NaAs and $\rm{Fe_{1-\textit{x}}Co_\textit{x}As}$, which was loaded into a small alumina crucible and sealed in a Nb tube. The Nb tubes were then sealed under vacuum in quartz tubes, which were heated at $\rm{950^\circ}$C for 24 h, cooled at $\rm{3^\circ}$C/h to $\rm{600^\circ}$C, and then quickly cooled to room temperature. After opening the tubes, the products were soaked in ethanol under a $\rm{N_2}$ environment to dissolve the excess NaAs flux. Well-formed square plate crystals of $\rm{NaFe_{1-\textit{x}}Co_{\textit{x}}As}$ were isolated.   

Single crystals of NaFeAs were also grown directly from a stoichiometric melt. Initially, polycrystalline NaFeAs was prepared by annealing an elemental mixture for 48 h each at temperatures of $\rm{775^\circ}$C and $\rm{800^\circ}$C, with an intermittent grinding. Following the second annealing step, phase purity was confirmed by powder X-ray diffraction.  The polycrystalline NaFeAs was loaded into a small alumina crucible, and subsequently sealed inside a Nb tube. The Nb tube was placed in a RF induction furnace under flowing $\rm{N_2}$ (to prevent oxidation of the Nb tube and its contents), where it was quickly heated to $\rm{1200^\circ}$C, soaked for 1 h, and then quickly cooled to room temperature. A melted ingot was recovered from the alumina crucible. The ingot was broken apart to reveal single crystalline fragments.  

The $\rm{NaFe_{1-\textit{x}}Co_{\textit{x}}As}$ crystals are 100-150 $\mu$m in size and quite brittle. Handling them with tweezers often causes them to crumble. These factors, along with their reactiveness in air makes obtaining good Andreev reflection spectra from them very challenging. Soft point contact junctions are formed on freshly cleaved \textit{c}-axis crystal surfaces as described earlier \cite{Arham} and $dI/dV$ across each junction is measured using a standard four-probe lock-in technique. 

$\rm{NaFeAs}$ has an antiferromagnetic ground state. Bulk superconductivity is achieved upon Co doping. \cite{Wang} We present results on underdoped $\rm{NaFe_{0.98}Co_{0.02}As}$ ($T_c$ $\sim$ 22.5 K) and overdoped $\rm{NaFe_{0.94}Co_{0.06}As}$ ($T_c$ $\sim$ 20.2 K).

\subsection{$\rm{NaFe_{0.98}Co_{0.02}As}$}

Figure 1 shows $dI/dV$ spectra for two different junctions on $\rm{NaFe_{0.98}Co_{0.02}As}$. The curves for the two junctions are plotted on the same bias voltage scale for comparison. The lowest temperature curves for both junctions (blue curves in Figures 1 (a) and 1 (c)) are picking up clear signals of Andreev reflection. Above $T_c$, the Andreev reflection dies out leaving behind a broad asymmetric conductance enhancement centered at zero bias voltage (red curves in Figures 1 (a) and 1 (c)). In fact, features corresponding to this enhancement coexist with superconductivity and are visible in the blue curves as well. The black arrows in the figures point them out. This has been seen in other iron pnictide and iron chalcogenide superconductors \cite{Arham} and a theoretical explanation as to why the conductance enhancement remains in the static nematic state is given in Ref [9]. Figures 1 (b) and 1 (d) show how this enhancement evolves with temperature for the junctions introduced in Figures 1 (a) and (c), respectively. For (b), the spectra have been normalized with respect to the value at -150 mV while for (d), they have been normalized to the value at -70 mV. With increasing temperature, the conductance enhancement is reduced. For (b), the enhancement disappears between 117 K and 151 K, leaving behind a concave up, weakly parabolic background. The junction in (d) is only biased up to $\pm$70 mV. The background conductance here is concave down throughout the measured temperature range, and is qualitatively similar to the 80 K curve in (b). For both junctions, the background is asymmetric.

The bulk resistivity of $\rm{NaFe_{0.98}Co_{0.02}As}$ is shown in the inset of Figure 1 (c). The resistivity decreases smoothly with falling temperature with no slope change that might correspond to a structural or magnetic transition.  

\begin{figure*}[htbp]
\centering
		\includegraphics[scale=0.7]{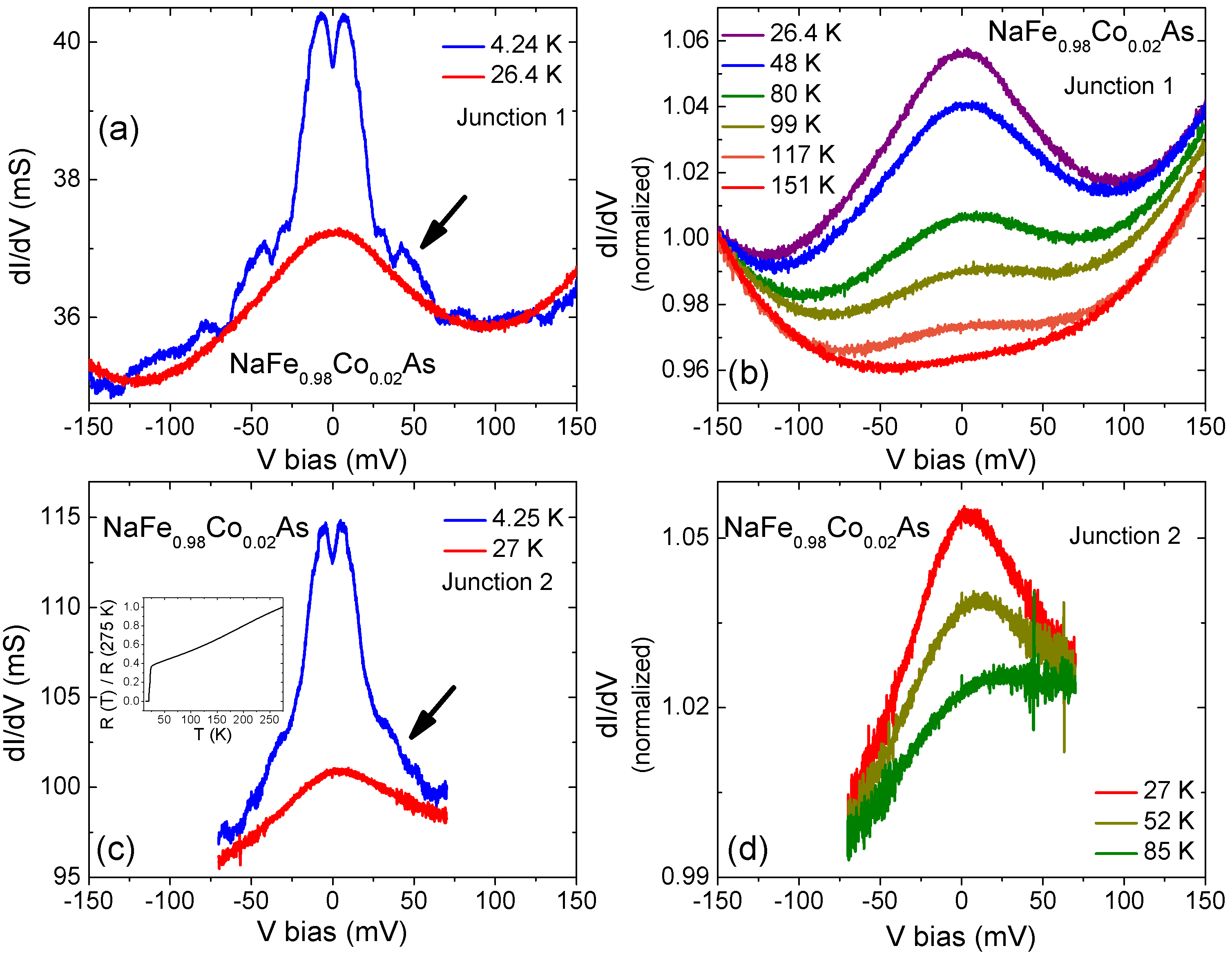}
	\caption{PCS $dI/dV$ spectra for two junctions on underdoped $\rm{NaFe_{0.98}Co_{0.02}As}$ ($T_c$ $\sim$ 22.5 K). The curves are plotted on the same bias voltage scale for comparison. (a, c) The lowest temperature curves for both junctions show features corresponding to Andreev reflection (blue curves). Above $T_c$, the Andreev reflection dies out leaving behind a broad asymmetric conductance enhancement centered at zero bias voltage (red curves). The arrows are pointing out that this enhancement coexists with the Andreev spectra at the lowest temperature. The inset to (c) shows the temperature dependence of the normalized bulk resistivity of the crystal. (b, d) The temperature evolution of the conductance enhancement around zero bias. Junction 1 is biased to $\pm$150 mV, and the enhancement disappears between 117 K and 151 K. Junction 2 is only biased up to $\pm$70 mV. At 85 K, the $dI/dV$ curve is still concave down and is qualitatively similar to the 80 K curve for Junction 1. For both junctions, the background is asymmetric.}
	\label{fig:100}
\end{figure*}  

Figure 2 shows the BTK fits to the low temperature Andreev spectra of the two junctions from Figure 1. The low temperature curves are normalized with those above $T_c$ to remove the background. The observed Andreev enhancement is very small for both junctions, 7.3$\%$ for junction 1 and 9.6$\%$ for junction 2. To fit these spectra, we assume that, along with normal metal-superconductor (N-S) transport channels, our junction also has parallel normal metal-normal metal (N-N) transport channels. This may result from part of the contact being non-superconducting due to contamination of the crystal surface on exposure to air. It is also possible that the Co doping is inhomogeneous and the full volume of the crystal is not superconducting, also giving rise to parallel N-S and N-N channels. This could also arise from a band not participating in the superconductivity, or that the Andreev coupling to some band(s) are weaker than others. We also point out that such a parallel channel model has been applied to heavy fermion superconductors, which are also multiband materials. \cite{WKPark} 

Our measured conductance may be described by the equation: 

\begin{equation}
\frac{dI}{dV}_{total} = w*\frac{dI}{dV}_{1:S} + \frac{dI}{dV}_{2:NS} 
\end{equation}
where S and NS represents the conductance arising from the superconducting (Andreev) and non-superconducting channels, respectively. We assume the non-superconducting term to be constant for all bias voltage. The fraction of the point contact that participates in Andreev reflection is denoted by \textit{w}.

Figure 2 (a) shows the BTK fit for junction 1 assuming two independent s-wave gaps (red solid curve). The gap values are $\rm{\Delta_1 = 5.0}$ meV and $\rm{\Delta_2 = 12.0}$ meV with \textit{w} = 0.5, meaning that 50$\%$ of the transport channels are N-S, with the rest being N-N. The value of \textit{w} was chosen to keep the broadening parameter $\Gamma\leq\Delta/2$ while maintaining a good fit. All the parameters for the fit are given in Table 1.     

It is worth attempting to simulate the experimental data by keeping \textit{w} = 1 and instead increasing the value of $\Gamma$, to reduce the Andreev enhancement. We find that a higher $\Gamma$ produces more broadened curves and cannot reproduce the sharp features that are observed experimentally. In addition, $\Gamma$ must increase, approaching $\Delta$ in value, which is an unphysical scenario.  
   
Figure 2 (b) shows two BTK fits for junction 2. Both of them assume two independent s-wave gaps but one of them is for \textit{w} = 1 (blue solid curve) and the other one is for \textit{w} = 0.3 (red dashed curve). The gap values for \textit{w} = 1 are $\rm{\Delta_1 = 5.0}$ meV and $\rm{\Delta_2 = 9.0}$ meV with $\Gamma_1/\Delta_1=0.76$ and $\Gamma_2/\Delta_2=0.7$. The gap values for \textit{w} = 0.3 are $\rm{\Delta_1 = 5.0}$ meV and $\rm{\Delta_2 = 11.0}$ meV. The value of \textit{w} was chosen to keep the ratio $\Gamma/\Delta$ between 0.2 and 0.3. All the parameters for the fits are given in Table 1. Notice that the red dashed curve tracks the low bias Andreev peaks while the blue solid curve overshoots them.     

\begin{figure*}[htbp]
\centering
		\includegraphics[scale=0.7]{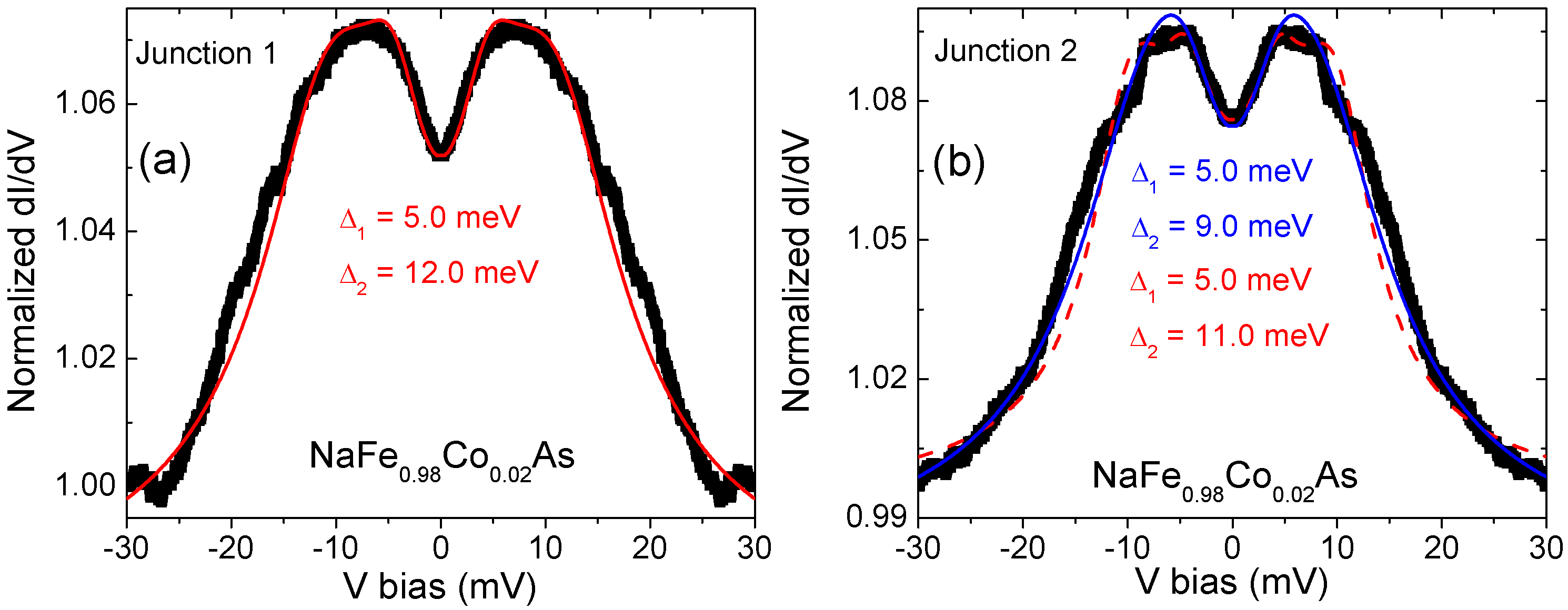}
	\caption{The Andreev spectra for $\rm{NaFe_{0.98}Co_{0.02}As}$ is fit to the BTK model using two independent s-wave gaps. The low temperature curves for junctions 1 and 2 from Figure 1 have been normalized with the curves above $T_c$ to remove the background. All the parameters for the fits are given in Table 1. (a) The fit for junction 1 (red solid curve) is for gaps with values $\rm{\Delta_1 = 5.0}$ meV and $\rm{\Delta_2 = 12.0}$ meV. Using Eqn(1), \textit{w} = 0.5, meaning that 50$\%$ of the transport channels are N-S, with the rest being N-N. The value of \textit{w} was chosen to keep the broadening parameter $\Gamma\sim\Delta/2$ while maintaining a good fit. (b) Two BTK fits for junction 2. One of them is for \textit{w} = 1 (blue solid curve) and the other one is for \textit{w} = 0.3 (red dashed curve). The gap values for \textit{w} = 1 are $\rm{\Delta_1 = 5.0}$ meV and $\rm{\Delta_2 = 9.0}$ meV with $\Gamma_1/\Delta_1=0.76$ and $\Gamma_2/\Delta_2=0.7$. The gap values for \textit{w} = 0.3 are $\rm{\Delta_1 = 5.0}$ meV and $\rm{\Delta_2 = 11.0}$ meV. The value of \textit{w} was chosen to keep the ratio $\Gamma/\Delta$ between 0.2-0.3. Notice that the red dashed curve tracks the low bias Andreev peaks while the blue solid curve overshoots them.}
	\label{fig:100}
\end{figure*}  

Some of our junctions on $\rm{NaFe_{0.98}Co_{0.02}As}$ do not show Andreev reflection below $T_c$, but rather a peculiar asymmetric feature that we reproduce in Figure 4 (a). For the two junctions shown, $dI/dV$ is larger for positive bias values than for the negative bias values. The gradient of the curve changes twice, at $\sim$ +5 mV and $\sim$ -5 mV. With increasing temperature, these features become thermally broadened and are not observable (not shown in figure). 

\subsection{$\rm{NaFe_{0.94}Co_{0.06}As}$}

$\rm{NaFe_{0.94}Co_{0.06}As}$ is overdoped with $T_c$ $\sim$ 20.2 K. The inset to Figure 3 (a) shows the bulk resistivity of a crystal. The resistivity does not exhibit any features corresponding to a structural or magnetic transition. 

Figures 3 (a), (c), and (d) show $dI/dV$ spectra for three different junctions on $\rm{NaFe_{0.94}Co_{0.06}As}$. They are plotted on the same bias voltage scale for comparison. 

The lowest temperature curves for all three junctions (blue curves in Figures 3 (a), (c), and (d)) show clear signals of Andreev reflection. The curves for junctions 2 and 3 are broadened at low biases. The inset to (c) shows that the $dI/dV$ curve is flat between $\pm$2 mV for junction 2 while the left inset to (d) shows that the $dI/dV$ curve is flat between $\pm$10 mV for junction 3. This could happen because of thermal population effects and the junctions being impacted by some inelastic scattering at the interface (diffusive regime). Junction 1 on the other hand shows sharp features at low voltage biases. Therefore, we perform BTK fitting on the Andreev spectrum of junction 1. 

Figure 3 (b) shows the junction 1 data symmetrized and normalized to the $dI/dV$ at 20 mV. (The junction resistance changed upon warming up so we are unable to normalize low temperature $dI/dV$ with the curve above $T_c$, as was done for the Andreev spectra in Figure 2.) Features corresponding to two superconducting gaps are observed, the arrows in the figure point them out. The BTK fit is done for two isotropic s-wave gaps. Using Equation 1 with \textit{w} = 0.28, we obtain $\rm{\Delta_1 = 4.95}$ meV and $\rm{\Delta_2 = 6.90}$ meV. All the parameters for the fit are given in Table 1. 

As mentioned earlier, for $\rm{NaFe_{0.98}Co_{0.02}As}$, a broad conductance enhancement around zero bias is observed above $T_c$ and this enhancement also coexists with the Andreev spectra below $T_c$. Below we show results of probing the spectra of $\rm{NaFe_{0.94}Co_{0.06}As}$ to see if such a feature is present or not.

At low temperatures, all three $\rm{NaFe_{0.94}Co_{0.06}As}$ junctions show some high bias features, close to 50 mV. The black arrows in Figure 3 (a), (c), and (d) point them out. For junctions 2 and 3, once the temperature is larger than $T_c$ (red curves in (c) and (d)), these features die out, leaving behind a V-shaped background. (For junction 1 we do not have spectra above $T_c$.) At $\rm{T = 22}$ K for junction 3, there is a slight dip in the $dI/dV$ around zero bias voltage, running from -15 mV to +15 mV. The right inset in Figure 3 (d) shows that this dip has filled up by 40 K, and further increase in temperature to 82 K causes no change in the spectra. 

\begin{figure*}[htbp]
\centering
		\includegraphics[scale=0.7]{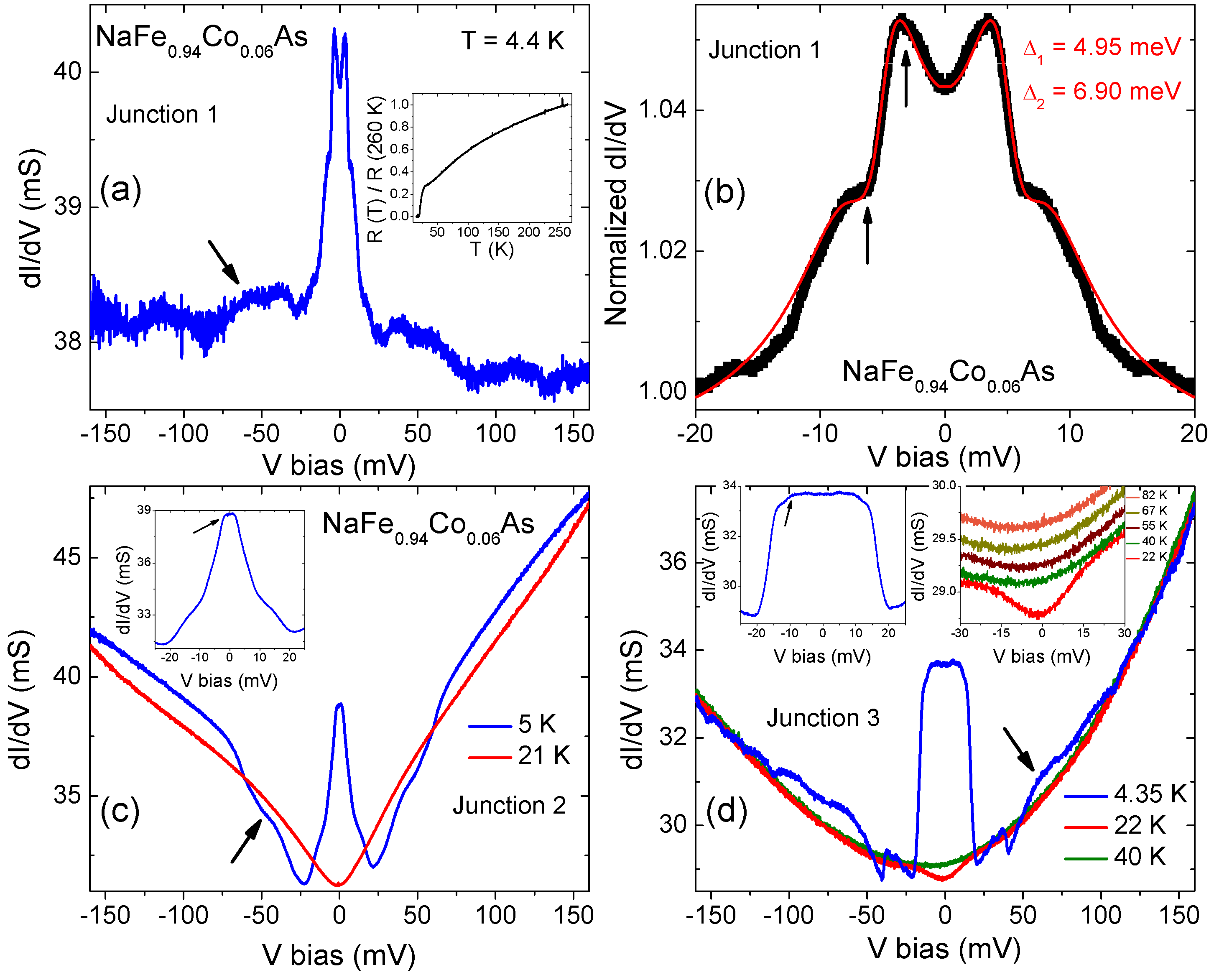}
	\caption{PCS $dI/dV$ spectra for three junctions on overdoped $\rm{NaFe_{0.94}Co_{0.06}As}$ ($T_c$ $\sim$ 20.2 K). (a) Junction 1 shows sharp features corresponding to Andreev reflection at low temperature. The arrow points out higher bias features close to 50 mV. The inset shows the bulk resistivity of the compound. (b) The Andreev spectra of junction 1 is fit to the BTK model (red curve) using two independent s-wave gaps with values $\rm{\Delta_1 = 4.95}$ meV and $\rm{\Delta_2 = 6.90}$ meV. The arrows in the figure point out the features corresponding to the two gaps. (c) Junction 2 detects Andreev spectra below $T_c$ (blue curve) along with a high bias feature that the arrow in the figure points out. Above $T_c$, only a V-shaped background remains (red curve). The inset shows that the Andreev spectra for junction 2 is smeared and $dI/dV$ is flat between $\pm$2 mV. (d) Junction 3 exhibits Andreev spectra below $T_c$ (blue curve) and a high bias feature that the arrow in the figure points out. Above $T_c$, a V-shaped background remains with a dip running for -15 mV to +15 mV (red curve). The right inset shows that this dip fills up by 40 K and further increase in temperature causes no change in the spectra. The left inset shows that the Andreev spectra for junction 3 is also smeared and $dI/dV$ is flat between $\pm$10 mV.}
	\label{fig:100}
\end{figure*}  

As we mentioned earlier for $\rm{NaFe_{0.98}Co_{0.02}As}$, some of our junctions on $\rm{NaFe_{0.94}Co_{0.06}As}$ also do not show Andreev reflection below $T_c$, but rather a peculiar asymmetric curve. Figure 4 (b) depicts such a spectra for two different junctions on  $\rm{NaFe_{0.94}Co_{0.06}As}$. These curves look fairly similar to the ones observed in $\rm{NaFe_{0.98}Co_{0.02}As}$ (Figure 4 (a)).

\begin{table*}[h]
\caption{BTK Fit Parameters $\rm{NaFe_{1-\textit{x}}Co_\textit{x}As}$}
\centering
 \begin{tabular}{c c c c c c c c c c c c}
 Doping & $\Delta_1$, $\Delta_2$ meV & $Z_1$, $Z_2$ & $\Gamma_1/\Delta_1$, $\Gamma_2/\Delta_2$ & $w=w_1+w_2$\\
 \hline
 2$\%$ Co, Junction 1 (red solid curve)   & 5.0, 12.0 & 0.47, 0.45 & 0.5, 0.46 & 0.5=0.15+0.35\\
 2$\%$ Co, Junction 2 (blue solid curve)	& 5.0, 9.0   & 0.43, 0.5 & 0.76, 0.7 & 1=0.4+0.6\\
 2$\%$ Co, Junction 2 (red dashed curve)	& 5.0, 11.0   & 0.39, 0.39 & 0.28, 0.23 & 0.3=0.09+0.21\\
 6$\%$ Co, Junction 1 (red solid curve)		& 4.95, 6.90   & 0.1, 1.5 & 0.12, 0.49 & 0.28=0.14+0.14\\ 
 \end{tabular}
\end{table*}

\begin{figure*}[htbp]
\centering
		\includegraphics[scale=0.7]{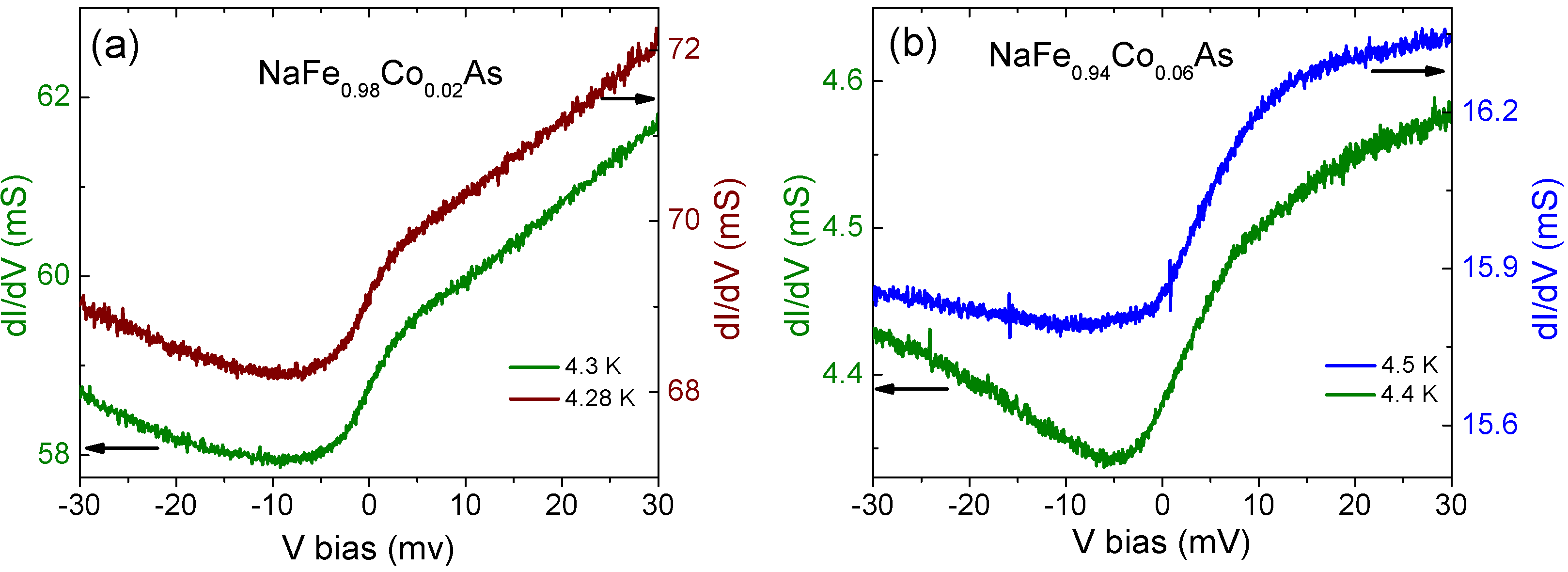}
	\caption{$dI/dV$ curves below $T_c$ that do not show Andreev reflection. The arrows in the figures are pointing out the y-axis corresponding to each curve. (a) Two different junctions constructed on $\rm{NaFe_{0.98}Co_{0.02}As}$. Instead of Andreev reflection, we observe a peculiar asymmetric feature. $dI/dV$ is larger for positive bias values than for the negative bias values. The gradient of the curve changes twice, at $\sim$ +5 mV and $\sim$ -5 mV. (b) Two different junctions for $\rm{NaFe_{0.94}Co_{0.06}As}$. These $dI/dV$ curves are similar to the ones obtained for 2$\%$ Co doping and shown in (a).}
	\label{fig:100}
\end{figure*}  

\subsection{$\rm{NaFeAs}$}

In the literature, $\rm{NaFeAs}$ is reported to have a structural phase transition at $\sim$ 50 K and a magnetic phase transition at $\sim$ 40 K. \cite{Wang} Completely pure $\rm{NaFeAs}$ does not superconduct. However, on exposure to air, oxidation occurs causing partial superconductivity. \cite{111ANL} Oxidizing the crystal gently with water extracts electrons and $\rm{Na^+}$ cations from the structure, yielding $\rm{Na_{1-\textit{x}}FeAs}$ with a maximum $T_c$ of $\sim$ 25 K. Oxidizing the sample more vigorously by exposure to air changes the structure to $\rm{NaFe_2As_2}$ ($\rm{ThCr_2Si_2}$-type) and results in a maximum $T_c$ of $\sim$ 12 K.  

We probe $\rm{NaFeAs}$ crystals grown from the melt (Figure 5 (a)) and from NaAs flux (Figure 5 (b)). They exhibit remarkably different spectra. For the melt-grown crystal, at the lowest temperature we detect a very weak Andreev signal. This signal is superimposed on a broad conductance enhancement. With increasing temperature the Andreev signal disappears, and the conductance peaks at $\sim$ 22 mV remain with a minimum developing at zero bias. As the temperature is further increased, the peaks move to lower bias and the conductance enhancement is reduced. The $dI/dV$ curve becomes completely flat around 90 K. The $dI/dV$ values in Figure 5 (a) have been normalized to the conductance at -200 mV. This spectra is reminiscent of the $dI/dV$ curves observed on other iron pnictide and chalcogenide parent compounds: $\rm{BaFe_2As_2}$, $\rm{SrFe_2As_2}$, $\rm{CaFe_2As_2}$, and $\rm{Fe_{1+\textit{y}}Te}$. \cite{Arham}

The situation for the flux-grown crystal is completely different. At the lowest temperature, $dI/dV$ develops a sharp dip at zero bias voltage. As the temperature is increased, the dip gets shallower and shallower, and disappears at $\sim$ 40 K, as pointed out by the black arrow in Figure 5 (b). This is also the temperature at which the antiferromagnetic transition occurs in the crystal. Any further increase in temperature does not change the spectra. The inset in the figure shows the curve obtained at 99 K. It is strongly asymmetric with the positive voltage bias showing higher conductance values than the negative voltage bias. The $dI/dV$ values in Figure 5 (b) have been normalized to the values at -100 mV, and all curves after the one at 4.3 K have been shifted vertically up by 0.005.

\begin{figure*}[thbp]
\centering
		\includegraphics[scale=0.7]{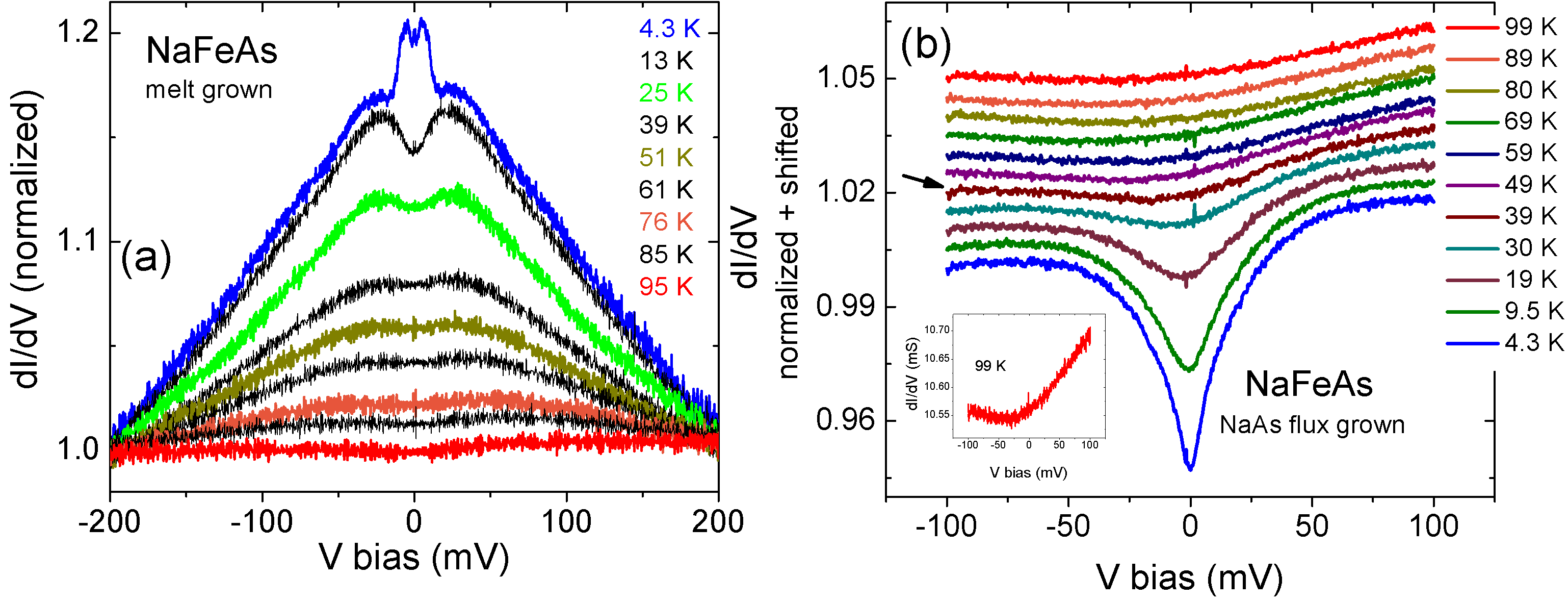}
	\caption{$dI/dV$ for $\rm{NaFeAs}$. (a) For melt-grown $\rm{NaFeAs}$, a weak Andreev signal is detected at the lowest temperature. A broad zero bias conductance enhancement is observed with a dip at the center. With increasing temperature, the enhancement is reduced and disappears around 90 K. (b) For flux-grown $\rm{NaFeAs}$, at the lowest temperature, $dI/dV$ develops a sharp dip at zero bias voltage. The dip disappears close to $T_N$ ($\sim$ 40 K), as pointed out by the black arrow in figure. Any further increase in temperature does not change the spectra. The inset shows $dI/dV$ for 99 K. It is strongly asymmetric with the positive voltage bias showing higher conductance values than the negative voltage bias.} 
	\label{fig:100}
\end{figure*}

\section{Discussion}

Certain iron-based superconductors exhibit an electronic nematicity that is reflected as an in-plane stress-induced resistive anisotropy above the structural phase transition in their detwinned normal state. \cite{Fisher} Our previous experiments have shown that PCS detects a conductance enhancement in the normal state of such iron based compounds (underdoped $\rm{Ba(Fe_{1-\textit{x}}Co_\textit{x})_2As_2}$, $\rm{Fe_{1+\textit{y}}Te}$, $\rm{SrFe_2As_2}$). We argue that orbital fluctuations in these compounds give rise to a non-Fermi liquid behavior and cause the conductance enhancement. \cite{Arham, Weicheng, OurReview, WeichengUnpublished} 

For underdoped $\rm{NaFe_{0.98}Co_{0.02}As}$ we detect a $dI/dV$ enhancement in the normal state. This enhancement coexists with the Andreev spectra below $T_c$. For overdoped $\rm{NaFe_{0.94}Co_{0.06}As}$ such a signal is not present. Underdoped $\rm{NaFe_{1-\textit{x}}Co_{\textit{x}}As}$ has a stress-induced in-plane resistive anisotropy above $T_S$. \cite{Na111R, Jiang} This matches up with the trend that PCS detects a conductance enhancement in the normal state if an in-plane stress-induced resistive anisotropy exists above the structural phase transition. Based on our PCS results, overdoped $\rm{NaFe_{1-\textit{x}}Co_{\textit{x}}As}$ does not exhibit an electronic nematic phase in its normal state.

Angle resolved photoemission spectroscopy (ARPES) on $\rm{NaFe_{1-\textit{x}}Co_{\textit{x}}As}$ detects nearly isotropic s-wave gaps of magnitude 6.5 meV and 6.8 meV in $x = 0.05$ ($T_c$ $\sim$ 18 K). \cite{ARPES111} Another ARPES experiment detects anisotropic gaps, varying between 4 and 7 meV on $\textit{x} = 0.0175$ ($T_c$ $\sim$ 18 K). \cite{ARPES2} Scanning tunneling microscopy (STM) detects gaps of size 5.5 meV on $\textit{x} = 0.028$ ($T_c$ $\sim$ 20 K) and 4.7 meV on $x = 0.061$ ($T_c$ $\sim$ 13 K). \cite{STM111} To our knowledge, there are no other reported results for point contact or tunneling spectroscopies on $\rm{NaFe_{1-\textit{x}}Co_{\textit{x}}As}$. 

The BTK fits to our PCS data are shown in Figures 2 and 3 (b). We extract gap values of 4.95 meV and 6.90 meV on $\textit{x} = 0.06$ ($T_c$ $\sim$ 20.2 K). These numbers are in good agreement with the gaps detected by ARPES and STM. For $\textit{x} = 0.02$ ($T_c$ $\sim$ 22.5 K), we observe one gap of magnitude 5.0 meV and a second gap of magnitude 11-12 meV. The second gap is significantly larger than the values detected by ARPES and STM. We speculate that the conductance enhancement that coexists with the Andreev reflection signal for $\textit{x} = 0.02$ has the effect of moving the Andreev spectra to higher biases, resulting in enlarged gap values. Figures 1 (a, c) show that the excess conductance enhancements at low temperatures and above $T_c$ are not identical, i.e. the blue and red curves do not overlap for $V>>\Delta$. Thus normalizing the low temperature spectra with the data above $T_c$ does not remove all the excess conductance enhancement due to orbital fluctuations. This likely contributes to the BTK fit giving the artificially large gap value.     

Figure 4 shows that instead of Andreev reflection below $T_c$, we occasionally pick up an anomalous, highly anisotropic $dI/dV$ signal from both $\rm{NaFe_{0.98}Co_{0.02}As}$ and  $\rm{NaFe_{0.94}Co_{0.06}As}$. A comparison with a recent STM paper helps in providing an explanation. \cite{Na111Hoffman} Song et al. show that while the surface of cleaved $\rm{Sr_{0.75}K_{0.25}Fe_2As_2}$ is dominated by the Sr/K layer, patches of As interspersed between the Sr/K layer also exist. The superconducting gap is only detected on the Sr/K layer, while the As patches show a gapless, anisotropic $dI/dV$ signal (Figure 1 (e) in Ref[21]) that is very similar to our Figure 4. It is conceivable that the surface of cleaved $\rm{NaFe_{1-\textit{x}}Co_{\textit{x}}As}$ is dominated by either Na or As layers, and we pick up Andreev reflection from the Na portions and the anomalous, anisotropic signal from the As patches.   

Like $\rm{NaFe_{0.98}Co_{0.02}As}$, melt-grown $\rm{NaFeAs}$ shows a conductance enhancement in the normal state reminiscent of what we previously observed on the 122 parent compounds and $\rm{Fe_{1+\textit{y}}Te}$. \cite{Arham} In addition, an in-plane resistive anisotropy that sets in above the structural transition has also been detected in $\rm{NaFeAs}$. \cite{Jiang} Thus it is likely that the same mechanism is at play in all these compounds and the conductance enhancement observed in $\rm{NaFeAs}$ is also a consequence of orbital fluctuations. 

The question remains why the spectra obtained from flux-grown $\rm{NaFeAs}$ are so different than that of the melt-grown. Instead of an enhancement, a dip develops in the conductance which disappears above $T_N$. STM $dI/dV$ shows a similar feature from $\rm{NaFeAs}$. \cite{NaFeAsSTM} The authors attribute it to the gapping of the the Fermi surface due to the spin density wave transition. In addition, both STM and PCS detect a similar shaped asymmetric background for $T>T_N$. 

As mentioned earlier, oxidation changes $\rm{NaFeAs}$ into $\rm{Na_{1-\textit{x}}FeAs}$ or $\rm{NaFe_2As_2}$. Our crystals are most likely a combination of all three structures. Different levels of purity in the melt-grown and flux-grown $\rm{NaFeAs}$ may cause the variance in our spectra. 

We also note that PCS spectra similar to the `V-shaped' curve obtained from flux grown $\rm{NaFeAs}$ have previously been observed on a variety of materials by our research group, and might be caused by disorder in the system. \cite{Altshuler, Gershenzon} In such a scenario, our data on flux-grown $\rm{NaFeAs}$ (Figure 5 (b)) does not reflect the intrinsic properties of the crystal.   
  
\section{Summary and Conclusions}

We use point contact spectroscopy to study the iron-based superconductor $\rm{NaFe_{1-\textit{x}}Co_{\textit{x}}As}$ with $\rm{\textit{x} = 0, 0.02, 0.06}$. Melt-grown $\rm{NaFeAs}$ and underdoped $\rm{NaFe_{0.98}Co_{0.02}As}$ show a broad conductance enhancement around zero bias voltage that coexists with the Andreev reflection and survives well into the normal state. This enhancement is not present for overdoped $\rm{NaFe_{0.94}Co_{0.06}As}$. Such a signal has previously been detected by PCS in certain iron-based superconductors and attributed to orbital fluctuations in the normal state giving rise to a non-Fermi liquid behavior. \cite{Arham} Thus our data provides evidence for the presence of electronic nematicity arising from orbital fluctuations in the normal state of $\rm{NaFeAs}$ and $\rm{NaFe_{0.98}Co_{0.02}As}$. 
 
The Andreev spectra for $\rm{\textit{x} = 0.02, 0.06}$ provides evidence for multiple superconducting gaps. We fit our lowest temperature data using the extended BTK model with two s-wave superconducting gaps. 
 
\begin{acknowledgments}

This work is supported as part of the Center for Emergent Superconductivity, an Energy Frontier Research Center funded by the US Department of Energy, Office of Science, Office of Basic Energy Sciences under Award No. DE-AC0298CH1088.

\end{acknowledgments}

\bibliography{myrefs}
\end{document}